\documentclass[twocolumn,prb,aps]{revtex4}
\usepackage{graphicx}

\newcommand{\ep}{\varepsilon}
\newcommand{\ga}{\gamma}

\newcommand{\be}{\begin{equation}}
\newcommand{\ee}{\end{equation}}
\newcommand{\ba}{\begin{eqnarray}}
\newcommand{\ea}{\end{eqnarray}}

\newcommand{\nn}{\nonumber}
\newcommand{\la}{\label} 
\newcommand{\w}{\omega} 
\newcommand{\hT}{\hat{T}}
\newcommand{\hV}{\hat{V}}
\newcommand{\hH}{\hat{H}}
\newcommand{\htH}{\hat{H}_A}
\newcommand{\pa}{\partial}
\def\oneone{\rlap 1\mkern4mu{\rm l}}
\def\t1{e_{_T}}
\def\v1{e_{_V}}
\def\ct{e_{_{TTV}}}
\def\cv{e_{_{VTV}}}
\def\tt{e_{_{TTTTV}}}
\def\tv{e_{_{VTTTV}}}
\def\vt{e_{_{TTVTV}}}
\def\vv{e_{_{VTVTV}}}
\def\fr{_{FR}}
\def\fft#1#2{{#1 \over #2}}
\def\ft#1#2{{#1 \over #2}}

\begin{document}
%\tightenlines
\title{Exact evolution of time-reversible symplectic integrators
\\and their phase error for the harmonic oscillator}

\author{Siu A. Chin and Sante R. Scuro}

\affiliation{Department of Physics, Texas A\&M University,
College Station, TX 77843, USA}

%\date{\today}
\begin{abstract}
%%
%%%%%%%%%%%%%%%%%%%%%%%%%%%%%%%%%%%%%%%%%%%%%%%%%%%%%%%%%%%%%%%%%%%%%%%%
%%
%%  Abstract
%%
%%%%%%%%%%%%%%%%%%%%%%%%%%%%%%%%%%%%%%%%%%%%%%%%%%%%%%%%%%%%%%%%%%%%%%%%
%%

The evolution of any factorized time-reversible symplectic
integrators, when applied to the harmonic oscillator, can be exactly
solved in a closed form. The resulting modified Hamiltonians
demonstrate the convergence of the Lie series expansions. They are 
also less distorted than modified Hamiltonian of non-reversible algorithms.  
The analytical form for the modified angular frequency can be used
to assess the phase error of any time-reversible algorithm.

\end{abstract}
\maketitle

\section {Introduction}
Symplectic integrators\cite{hairer} are methods of choice for solving
diverse physical problems ranging from celestial mechanics\cite{yoshi},
molecular dynamics\cite{skeel}, to accelerator physics\cite{dragt88}
and lattice gauge theory\cite{creutz}. In contrast to other numerical 
methods for solving Hamiltonian dynamics, symplectic 
integrators evolves the system according to a modified
Hamiltonian\cite{dragt,yoshi} that can be made arbitrary close to 
the original Hamiltonian. For this reason, despite the fact that symplectic 
integrators can preserve all Poincar\'e invariants, they can never 
exactly conserve energy\cite{ge}. Otherwise, an integrator's 
modified Hamiltonian would have coincided with the original 
Hamiltonian, and the integrator's evolution would have been exact. 

Symplectic integrators can be derived most easily  
by the method of factorization/composition\cite{yoshi,mcl02}, 
where the algorithm is composed of many elementary ones with 
varying coefficients. In this case, one can systematically 
compute the modified Hamiltonian order-by-order via the
Lie series\cite{dragt} approach by using the  
Baker-Campbell-Hausdorff (BCH) expansion\cite{dragt,yoshi}. 
While this approach can demonstrate the existence of the  
modified Hamiltonian perturbatively, and show that it can be made 
arbitrary close to the original Hamiltonian, 
one has no sense of the modified Hamiltonian's global 
structure nor of the series' convergence.

In this work, we show that the evolution of {\it any} factorized,
time-reversible integrators, can be exactly solved for the 
harmonic oscillator. The algorithm's evolution
remained that of a harmonic oscillator but with modified 
mass and spring constant. Thus the model illustrates very
simply, how reversible symplectic algorithms preserve 
invariant tori of a dynamical system. Moreover, the closed form 
expression for the modified Hamiltonian, when expanded, matches 
the Lie series order by order, thus demonstrating 
the latter's convergence. Finally, we show that the analytical
modified oscillator frequency can be used to benchmark any 
time-reversible algorithm.            

In the following, we will briefly summarize key features of 
the Lie-Poisson construction of symplectic integrators, the Lie series 
expansion for the modified Hamiltonian, the phase-space matrix 
approach, and comparative results for a number of fourth order
reversible algorithms. 

\section {Symplectic integrators}
The evolution of any dynamical variable $W(q_i,p_i)$,
is given by the Poisson bracket, and therefore by the
corresponding Lie operator $\hat H$ associated with the Hamiltonian
function $H(q_i,p_i)$, {\it i.e.}
%%%%%
\ba
\fft{dW}{dt} &=&\{W,H\}\nn\\
             &=& \Bigl(\fft{\pa H}{\pa p_i}\fft{\pa}{\pa q_i}
         -\fft{\pa H}{\pa q_i}\fft{\pa}{\pa p_i}\Bigr)W
             = \hH W . \la{liou}
\ea
%%%%%
For any dynamical function $Q$, we can similarly define its 
associated Lie operator $\hat Q$ via the Poisson bracket
\be
\hat{Q} W=\{W,Q\}.
\la{liop}
\ee
%This is the fundamental equation for the Lie-Poisson construction
%of symplector integrator.
The operator equation (\ref{liou}) can be formally solved via
%%%%%
\be
W(t)={\rm e}^{t \hH}W(0) .
\la{formsol}
\ee
%%%%%
Symplectic algorithms are derived by approximating the evolution
operator ${\rm e}^{t \hH}$ for a short time in a product form.
For Hamiltonian function of the usual separable form,
%%%%%
\be
H({\bf q},{\bf p}) = T({\bf p})+V({\bf q}),\quad {\rm with}\quad
T({\bf p}) = \fft{1}{2}p_i p_i ,\la{ham}
\ee
%%%%%
the Hamiltonian operator is also separable,
%%%%%
%\be
$
\hH=\hT+\hV , \la{htv}
$
%\ee
%%%%%
with differential operators $\hat{T}$ and $\hat{V}$
given by
\be
\hT \equiv\{\cdot,T\}=\fft{\pa T}{\pa p_i}\fft{\pa}{\pa q_i}
=p_i \fft{\pa}{\pa q_i},
 \la{htop}
 \ee
 \be
\hV \equiv\{\cdot,V\}= -\fft{\pa V}{\pa q_i} \fft{\pa}{\pa p_i}
=F_i({\bf q}) \fft{\pa}{\pa p_i}.
\la{hvop}
\ee
The Lie transforms\cite{dragt} 
${\rm e}^{\ep \hT}$ and ${\rm e}^{\ep \hV}$,
are then displacement operators which shift $q_i$ and $p_i$ forward in
time via
%%%%%
\be
{\bf q}\rightarrow {\bf q}+\ep {\bf p}
\quad{\rm and}\quad {\bf p}\rightarrow {\bf p}+\ep {\bf F} .\la{pqsh}
\ee
%%%%%
Thus, if ${\rm e}^{\ep \hH}$ can be factorized into products of
Lie transforms ${\rm e}^{\ep \hT}$ and
${\rm e}^{\ep \hV}$, then each factorization gives rise to an integrator
for evolving the system forward in time. Existing literature on symplectic 
algorithms are concerned with decomposing ${\rm e}^{\ep \hH}$
to high orders in the product form
%%%%%
\be
{\rm e}^{\ep(\hT+\hV)}=\prod_{i=1}^N
{\rm e}^{t_i\ep \hT}{\rm e}^{v_i\ep \hV}, \la{prod}
\ee
%%%%%
with a set of well chosen factorization coefficients
$\{t_i,v_i\}$. In most cases, we will consider only
(left-right) symmetric factorization schemes 
such that  either $t_1=0$ and
$v_i=v_{N-i+1}$, $t_{i+1}=t_{N-i+1}$, or $v_N=0$ and
$v_i=v_{N-i}$, $t_{i}=t_{N-i+1}$. In either cases, the algorithm
is exactly {\it time-reversible}, and the energy error terms
can only be an even function of $\ep$.
Such a symmetric factorizations is then at least second
order. 

\section {Lie series expansion }

A well known symmetric algorithm is
the second order St\"ormer/Verlet (SV)  
algorithm,
\be
{\cal T}_{SV}(\ep)\equiv
{\rm e}^{\fft12\ep\hV}
{\rm e}^{\ep\hT}
{\rm e}^{\fft12\ep\hV}={\rm e}^{\ep\htH},
\la{tbopt}
\ee
where $\htH$ is the approximate Hamiltonian operator
of the form 
%%%%%
\ba
&&\htH = \hT+\hV
+\ep^2\left(\ct[\hT^2\hV]+\cv[\hV\hT\hV]\right)\nn \\
&&+\ep^4\left(\tt[\hT\hT^3\hV]
+ \tv[\hV\hT^3\hV]\right. \nn \\
&&\left. +\vt[\hT(\hT\hV)^2]
+\vv[\hV(\hT\hV)^2]\right)+..,
\la{hop4th}
\ea
%%%%%
where $\ct$, $\tv$ etc., are coefficients specific to a particular
algorithm. We have used the condensed commutator notation
$[\hT^2\hV]\equiv[\hT,[\hT,\hV]]$. For the case of algorithm SV,
the error coefficients are
\ba
\ct=\fft{1}{12},&&\cv=\fft{1}{24},\nn\\
\tt=-\fft1{720},&&\tv=\fft1{360},\nn\\
\vt=-\fft1{120},&&\vv=-\fft1{480}.
\la{errcof}
\ea 
These are computed by repeated applications of the  
Baker-Campbell-Hausdorff (BCH) formula for combining exponentials of
operators. Thus the algorithm evolves the
system according to $\htH$ rather than the true Hamiltonian $\hat{H}$.
Knowing $\htH$ then allows us to determine the actual Hamiltonian 
function ${H_A}$ which governs the algorithm's evolution. 
From the fundamental definition of Lie operator (\ref{liop}), one can
convert operators back to functions via $[T,V]\rightarrow\{V,T\}=-\{T,V\}$.
For symmetric factorization with only even order commutators, 
the Hamiltonian function can be obtained simply by 
replacing condensed commutators with condensed
Poisson brackets: $[\hT^2\hV]\rightarrow 
\{T^2 V\}\equiv\{T,\{T,V\}\}$, {\it etc.}. 
For the harmonic oscillator,
\be
H(q,p)=\fft12{p^2}+\fft12\,\w^2\,q^2\, ,\la{h}
\ee
the non-vanishing Poisson brackets are:
\ba 
\{TTV\}=\w^2 p^2,&&
\{VTV\}=-\w^4 q^2,\nn\\
\{T(TV)^2\}=-2\w^4 p^2,&&
\{V(TV)^2\}=2\w^6 q^2.
\la{hopb}
\ea
There is thus a clear separation between the error coefficients,
which are characteristics of the algorithm
and the Poisson brackets, which are properties of the underlying
Hamiltonian. If the SV algorithm is applied to the harmonic 
oscillator, then the resulting modified Hamiltonian function is
%%%%%
\be
H_A(q,p)=\fft1{2\,m^*}\,p^2+\fft12\,k^*\,q^2,
\la{2ndrha}
\ee
with effective mass and spring constant,
%%%%%
\ba
\fft1{m^*}&=& 1+\fft1{6}\ep^2\w^2
+\fft1{30}\ep^4\w^4+\cdots
     ,\la{calp}\\
\fft{k^*}{\w^2}&=&1-\fft1{12}\ep^2\w^2
-\fft1{120}\ep^4\w^4+\cdots
     ,\la{calq}
\ea
and approximate angular frequency $\w_A=\sqrt{\fft{k^*}{m^*}}$.
The algorithm therefore evolves the system according to
\be
\pmatrix{q(t) \cr p(t) }=
%\pmatrix{\cos(\w_A t) & (m^*\w_A)^{-1}\sin(\w_A t)\cr
%          -(m^*\w_A)\sin(\w_A t)& \cos(\w_A t) }
\pmatrix{\cos(\w_A t) & \fft{\sin(\w_A t)}{\sqrt{m^*k^*}}\cr
          -\sqrt{m^*k^*}\sin(\w_A t)& \cos(\w_A t) }
\pmatrix{q_0 \cr p_0}.
\la{solut}
\ee
By contrast, a first order, non-reversible factorization  
%%%%%
\be
{\rm e}^{\ep\hT} {\rm e}^{\ep\hV} = {\rm e}^{\ep\htH}
 ,\la{exprod}
\ee
%%%%%
with $\htH$ directly given by the BCH formula,
%%%%%
\be
\htH= \hH+\fft12\ep[\hT\hV]+\fft1{12}\ep[\hT\hT\hV]
   -\fft1{12}\ep[\hV\hT\hV]+\dots , \la{bch}
\ee
produces the following expansion,
\ba
{H_A} = \fft12[{p^2}+\w^2q^2+\ep\w^2p\,q&]&
\bigl( 1+\fft1{6}\ep^2\w^2\nn\\
&&+\fft1{30}\ep^4\w^4+\dots\bigr).
%+\fft12\ep\w^2pq
%+\fft1{12}\ep^2\w^2[p^2+\w^2 q^2]\nn\\
%&&+\fft1{12}\ep^3\w^4pq+
%\fft1{60}\ep^4\w^4[p^2+\w^2 q^2]\\
%&&+\fft1{60}\ep^5\w^6pq+\dots .\nn
\la{power}
\ea
Note that the overall factor is the same as (\ref{calp}).
The odd order terms distort the harmonic oscillator
fundamentally, producing a distinctive, 45$^\circ$-tilted 
phase-space ellipse, commonly seen in the standard map\cite{greene}, 
or when solving the pendulum\cite{yoshi}.
This is an artifact of not being time-reversible, a property 
foreign to the original dynamics. Time-reversible algorithms with
modified Hamiltonian such as (\ref{2ndrha}) showed no such 
distortion\cite{hair2}.

\section{Matrix method}
   
The above expansion method for computing the modified Hamiltonian,
while general, is severely limited to low orders.
At higher order, the number of commutator and Poisson brackets 
proliferates greatly due to the complexicity of the BCH expansion,
it is then difficult to assess the convergence
of series (\ref{calp}), (\ref{calq}) and (\ref{power}).  

To compute $H_A$ exactly, we completely abandon the BCH-based
approach. For the 1-D harmonic oscillator, let's denote
the phase-space vector at the $i$th iteration of the algorithm
as 
%%%%%
\be                                        
{\bf r}_i \equiv \pmatrix{q_i \cr p_i}.
\la{rmatr}
\ee
For the SV algorithm,
let ${\bf T}$ and ${\bf V}$ denote the effect of
Lie operators ${\rm e}^{\ep\hT}$ and
${\rm e}^{\fft12\ep\hV}$ acting on these vectors. 
From (\ref{pqsh}), it is easy to see that
they are upper and lower triangular matrices given by
%%%%%
\be
{\bf T}=\pmatrix{1 & \ep \cr 0 & 1 } ,\quad
{\bf V}=\pmatrix{1 & 0 \cr -\fft12\w^2\ep & 1 } .
\la{tvmat}
\ee
%%%%%
Hence, the algorithm corresponds to the 
product matrix,
%%%%%
\be
{\bf M}\equiv{\bf V}{\bf T}{\bf V}
=\pmatrix{1-\fft12{\ep^2\w^2} &\ep\cr
-\ep\w^2(1-\fft14{\ep^2\w^2})&
1-\fft12{\ep^2\w^2} }.\la{2ndmap}
\ee
%%%%%
Matrix representations have been widely used to study
the stability\cite{hardy,gray} and convergence\cite{skeel,mcl94}
of low order symmetric integrators. Our interest here is not
to solve the harmonic oscillator per se, but to use it as a mean
of obtaining the converged value of the Lie series. In contrast
to previous use of the matrix method to study the harmonic 
oscillator\cite{skeel,hardy,gray,mcl94}, we are not interested
how each algorithm's approximate solution approaches the 
exact solution; we are interested only in the exact form 
of the approximate solution itself. Below,
we also emphasize the distinction between time-reversible and
time-irreversible algorithms.
As illustrated in (\ref{tvmat}), for quadratic Hamiltonians,
each elemental operator
${\rm e}^{t_i\ep\hT}$ and ${\rm e}^{v_i\ep\hV_i}$
can in general be represented as upper or lower triangular matrices,
\be
{\bf T}_i(\ep)=\pmatrix{1&\sigma_i(\ep)\cr0&1},{\quad}
{\bf V}_i(\ep)=\pmatrix{1&0\cr-\mu_i(\ep)&1},
\ee
where the {\it constant} matrix elements 
$\sigma_i(\ep)$ and $\mu_i(\ep)$ are both positive
as $\ep\rightarrow 0$ but are {\it odd} functions of $\ep$:
\be
\sigma_i(-\ep)=-\sigma_i(\ep),{\quad} 
\mu_i(-\ep)=-\mu_i(\ep).
\la{oddft}
\ee
Both ${\bf T}_i$ and ${\bf V}_i$ can be further decomposed as
%%%%%
\ba
&&{\bf T}_i(\ep)=
\pmatrix{1&0\cr0&1}+\sigma_i(\ep)\pmatrix{0&1\cr0&0}
=\oneone + \sigma_i(\ep){\bf t},\qquad\la{matgent}\\
&&{\bf V}_i(\ep)=
\pmatrix{1&0\cr0&1}-\mu_i(\ep)\pmatrix{0&0\cr 1&0}
=\oneone -\mu_i(\ep){\bf v},\qquad\la{matgenv}
\ea
%%%%%
where ${\bf t}$ and ${\bf v}$ are nilpotent matrices,
\be
{\bf t}^2=0,{\quad} {\bf v}^2=0,
\la{nil}
\ee
with property
$
{\bf t}{\bf v}+{\bf v}{\bf t}=\oneone.
$
The nilpotence of ${\bf t}$ and ${\bf v}$
makes it obvious that both ${\bf T}_i$ and ${\bf V}_i$ are 
{\it time-reversible},
\be
 {\bf T}_i(-\ep){\bf T}_i(\ep)=\oneone,{\quad}
 {\bf V}_i(-\ep){\bf V}_i(\ep)=\oneone.
\ee
The time-reversibility of ${\bf T}_i(\ep)$ and 
${\bf V}_i(\ep)$ can also be traced to the fact that they have 
{\it equal} diagonal elements
{\it even} in $\ep$ and off-diagonal elements {\it odd} in $\ep$. 
In fact such a general matrix
\be
{\bf G}(\ep)=g(\ep^2)\oneone
+\tau(\ep){\bf t}-\nu(\ep){\bf v}
=\pmatrix{g &\tau\cr
-\nu&g}
\la{gentr}
\ee
with 
%\be
$\tau(-\ep)=-\tau(\ep)$ and 
$\nu(-\ep)=-\nu(\ep)$,
%\la{oddtn}
%\ee
is indeed time-reversible,
\ba
{\bf G}(-\ep){\bf G}(\ep)
&=&g^2\oneone-(\tau{\bf t}-\nu{\bf v})^2
\nn\\
 &=& (g^2+\nu\tau)\oneone=(\det{\bf G})\oneone,
\la{trev}
\ea
provided that $(\det{\bf G})=1$. Since this is
true of any symplectic integrator, we conclude
that any symplectic integrator of the form (\ref{gentr}),
is time-reversible. 
The generalization of (\ref{2ndmap}), corresponding to
the symmetric form of (\ref{prod}), can be now
written as 
%%%%%
\be
{\bf M}(\ep)=\prod_i{\bf T}_i(\ep){\bf V}_i(\ep).
\la{corrgen}
\ee
%%%%%
The advantage of the matrix representation is clear.
Each matrix product can be computed exactly, whereas the
corresponding product of exponential operators would 
have required an infinite BCH expansion.
%%%%%
For symmetric factorization, ${\bf M}$ is time-reversible. 
We now prove by
induction that ${\bf M}$ is indeed of the form (\ref{gentr}). 
Let ${\bf G}$ be of the form (\ref{gentr}), then
\ba
{\bf G}^\prime&=&{\bf T}_i{\bf G}{\bf T}_i,\nn\\
    &=&g{\bf T}_i^2+\tau{\bf T}_i{\bf t}{\bf T}_i
            -\nu{\bf T}_i{\bf v}{\bf T}_i,\nn\\
&=&(g-\nu\sigma_i)\oneone+(\tau+2g\sigma_i-\nu\sigma_i^2){\bf t}
-\nu{\bf v},
\la{protr}
\ea
which is again of the form (\ref{gentr}). 
Similarly for
${\bf V}_i{\bf G}{\bf V}_i$,
\ba
{\bf G}^\prime
    &=&g{\bf V}_i^2+\tau{\bf V}_i{\bf t}{\bf V}_i
            -\nu{\bf V}_i{\bf v}{\bf V}_i\nn\\
&=&(g-\tau\mu_i)\oneone+\tau{\bf t}
-(\nu+2 g\mu_i-\tau\mu_i^2){\bf v}.
\la{prots}
\ea
Since the initial ${\bf G}$ is either ${\bf T}_i$ or 
${\bf V}_i$, which is of the form (\ref{gentr}),
the proof is complete. If the algorithm is {\it not}
time-reversible, then the diagonal elements
will not be equal and {\bf M} will have the more general form
\be
{\bf M}=\pmatrix{g&\tau\cr -\nu&h}. \la{genG}
\ee  
We will consider this more general matrix 
and recover the time-reversible case at the end by
setting $h=g$.   

For the harmonic oscillator, because the matrix ${\bf M}$
is constant for all time steps, we have 
\be
{\bf r}_{_N} = {\bf M}^N{\bf r}_0.
\la{rsubn}
\ee
To evaluate ${\bf M}^N$ analytically, we diagonalize 
{\bf M} via 
%%%%%
\be
{\cal M}={\bf S}^{-1}{\bf M}{\bf S} .\la{Gsymtrasf}
\ee
For $\nu\tau>(g-h)^2/4$, the eigenvalues are unitary,
\be
\lambda_{\pm}=\fft{g+h}2\pm i\xi
={\rm e}^{\pm i {\theta}},\la{Geigen}
\ee
where 
\ba
\xi&=&\sqrt{\nu\tau-(g-h)^2/4} ,\la{sig}\\
{\theta}&=&\cos^{-1}\left(\fft12\,{\rm Tr}\,{\bf M}\right)
=\cos^{-1}\left(\fft{g+h}2\right).\la{Gangle}
\ea
The diagonal matrix and the transform are therefore
\be
{\cal M}=
\pmatrix{{\rm e}^{-i{\theta}}&0\cr 0&{\rm e}^{i{\theta}}},\quad
{\bf S}=\pmatrix{\fft{(h-g)/2+i\xi}{\nu}&
\fft{(h-g)/2-i\xi}{\nu}\cr1&1}.
\ee
The $N$th power of {\bf M} can now be computed
via 
$$
{\bf M}^N = {\bf S}{\cal M}^N{\bf S}^{-1}
= {\bf S}{\rm e}^{N\ln{\cal M}}{\bf S}^{-1}
= {\rm e}^{{\bf S}(N\ln{\cal M}){\bf S}^{-1}}
= {\rm e}^{\bf \Lambda} ,
$$
where we have substituted $N=t/\ep$,
\ba
{\bf \Lambda}&=&\fft{t}{\ep}(i{\theta})
{\bf S}\pmatrix{-1 &0\cr0& 1}{\bf S}^{-1}\\
&=&\fft{t\,\theta}{\ep\,\xi}
\pmatrix{\fft{g-h}2 & \tau \cr -\nu & -\fft{g-h}2}
=\pmatrix{\ga_0\,t&\ga_1\,t\cr -\ga_2\,t&-\ga_0\,t},
\la{GaMNgen}
\ea
yielding finally
\be
\ga_0=\fft{\theta}{\ep}\,\fft{(g-h)}{2\xi},\quad
\ga_1=\fft{\theta}{\ep}\,\fft{\tau}{\xi},\quad
\ga_2=\fft{\theta}{\ep}\,\fft{\nu}{\xi}.\la{Ggs}
\ee
%%%%%
The exponentiation of ${\bf \Lambda}$ then gives,
%%%%%
\be
{\bf M}^N={\bf R}+{\bf \Sigma}\, ,\la{RS}
\ee
where {\bf R} is a pure rotation and ${\bf \Sigma}$,
 a translation: 
\ba
{\bf R}&=&\pmatrix{\cos(\fft{\theta t}{\ep})&\fft{\tau}{\xi}
\sin(\fft{\theta t}{\ep})
\cr -\fft{\nu}{\xi} \sin(\fft{\theta t}{\ep})
&\cos(\fft{\theta t}{\ep})} \la{GRot}\\[.3cm]
{\bf \Sigma}&=&\ft{(g-h)}{2\xi} \sin\left(\ft{\theta t}{\ep}\right)
\pmatrix{1 &0\cr0& -1}.\la{GRefl}
\ea
%%%%%
For time-reversible algorithms, 
$h=g$, ${\bf \Sigma}=0$ and $\xi=\sqrt{\nu\tau}$.
By comparing (\ref{GRot}) to (\ref{solut}) one can quickly
identify
\be
{\w_A} =\fft{\theta}{\ep},\quad
\fft1{m^*}=\w_A\sqrt{\fft{\tau}{\nu}},\quad 
k^*=\w_A\sqrt{\fft{\nu}{\tau}},
\la{wapprox}
\ee
and the modified 
Hamiltonian conserved by the algorithm,
\be
{H_A}=\fft{\w_A}{2}\Bigl(\sqrt{\fft{\tau}{\nu}}p^2
+\sqrt{\fft{\nu}{\tau}}q^2\Bigr).
\la{haapprox}
\ee
The analytical form of the algorithm's solution (\ref{GRot}) 
is interesting in that the discrete character of the algorithm 
has disappeared! The solution is a continuous function of $t$, 
not necessarily a multiple of $\ep$.

For non-reversible algorithm such as (\ref{exprod}),
${\bf \Sigma}\neq 0$. This operator translates
$q\rightarrow q+s$ and $p\rightarrow p-s$ periodically,
thus causing an elongation of the phase-space ellipse
along the minor diagonal at -45$^\circ$. This is the
phase-space distortion alluded to earlier and observed
in the standard map\cite{greene}. While non-reversible 
algorithms may be of interest mathematically,
from the perspective of physics, there is nothing to be 
gain in solving time-reversible dynamics with non-reversible 
algorithms. 
  
\section{Modified Hamiltonians and the phase error}
To illustrate how this formalism works, for the 
SV algoritm, we have from (\ref{2ndmap}), 
\be
g=1-\fft12{\ep^2\w^2},\quad
\tau = \ep,\quad
\nu=\ep\w^2(1-\fft14\ep^2\w^2).
\la{gtn}
\ee 
From these we can determine in closed forms 
\ba
\w_A&=& \fft1{\ep}\cos^{-1}\bigl({1-\fft12\ep^2\w^2}
\bigr),\la{svwa}\\
\fft1{m^*}&=&\fft{\w_A}{\w}(1-\fft14\ep^2\w^2)^{-1/2},
\la{svm}\\
k*&=&\w_A{\w}(1-\fft14\ep^2\w^2)^{1/2}.
\la{svk}
\ea
We can now expand these physical quantities to any
order, but for the lack of space, we will stop at the
6th:
\ba
\fft{\w_A}{\w}&=&1
+\fft{\ep^2\w^2}{24}
+\fft{3\ep^4\w^4}{640}
+\fft{5\ep^6\w^6}{7168}
%+\fft{35\ep^8\w^8}{294912}
+\cdots,
\la{waexp}\\
\fft1{m^*}&=&1
+\fft{\ep^2\w^2}{6}
+\fft{\ep^4\w^4}{30}
+\fft{\ep^6\w^6}{140}
%+\fft{\ep^8\w^8}{630}%+\dots%\nn\\
%&&+\fft{\ep^{10}\w^{10}}{2772}
%+\fft{\ep^{12}\w^{12}}{12012}
+\cdots,
\la{calm}\\
\fft{k^*}{\w^2}&=&1
-\fft{\ep^2\w^2}{12}
-\fft{\ep^4\w^4}{120}
-\fft{\ep^6\w^6}{840}
%-\fft{\ep^8\w^8}{5040}%\nn\\
%&&-\fft{\ep^{10}\w^{10}}{27720}
%-\fft{\ep^{12}\w^{12}}{144144}
+\cdots .
\la{calg}
\ea
The first three terms in each case are in agreement with 
(\ref{calp}) and (\ref{calq}). 
By comparing (\ref{waexp}) to (\ref{svwa}) one can
immediately conclude that for $0<\ep\w<2$, the Lie series
converges. Thus the Lie series converges up to the
instability point. For another
example, in the case of the 
fourth order Forest-Ruth integrator\cite{forest},
(We use factorization coefficients as derived by Creutz and 
Gocksch\cite{creutz} and Yoshida\cite{yoshid}.)
\ba
g_{\fr}&=&1-\fft{\ep^2\w^2}{288}\left(144-12\ep^2\w^2-
(6+5\sqrt[3]{2}+4\sqrt[3]{2^2})\ep^4\w^4\right)\,,\nn\\
\tau_{\fr} &=& \ep \left(1-\fft{\ep^2\w^2}6-\fft{1+\sqrt[3]{2}\ep^4\w^4}
{72 \sqrt[3]{2^2}}\right.\,\nn\\
&&\left.+\fft1{1728}(25+20\sqrt[3]{2}+16\sqrt[3]{2^2})
\ep^6\w^6\right)\,,\nn\\
\nu_{\fr}&=&\ep\w^2\left(1-\fft{\ep^2\w^2}6-\fft1{144}
(4+4\sqrt[3]{2}+3\sqrt[3]{2^2})\ep^4\w^4\right)\,.\nn
\la{gtnfr}
\ea 
From these we can determine $\w_A$, $1/m*$ and $k^*$ via
(\ref{wapprox}).
They can then be expanded to any
order, and for lack of space, we will stop at the
6th:
\ba
\fft{\w_A}{\w}&=&1-\fft{(32+25\sqrt[3]{2}
+20\sqrt[3]{2^2})\ep^4\w^4}{1440}\nn\\
&&-\fft{(89+70\sqrt[3]{2}+56\sqrt[3]{2^2})\ep^6\w^6}{24192}
%&&-\fft{192+152\sqrt[3]{2}+121\sqrt[3]{2^2}\ep^8\w^8}{165888}\,\nn\\
%&&\left.-\fft{488+385\sqrt[3]{2}+308\sqrt[3]{2^2}\w^{10}\ep^{10}}{10948608}
+\cdots,\nn\\
\fft1{m^*}&=&1-\fft{(6+5\sqrt[3]{2}+5\sqrt[3]{2^2})\ep^4\w^4}{720}\,\nn\\
&&+\fft{(71+56\sqrt[3]{2}+42\sqrt[3]{2^2})\ep^6\w^6}{12096}
%+\fft{(29+23\sqrt[3]{2}+17\sqrt[3]{2^2})\ep^8\w^8}{51840}\,
%\nn\\
%&&+\fft{9701+7722\sqrt[3]{2}+5995\sqrt[3]{2^2}\w^{10}\ep^{10}}{47900160}\,\nn\\
%&&+\fft{37234+29471\sqrt[3]{2}+23608\sqrt[3]{2^2}\w^{12}\ep^{12}}{679311360}
+\cdots,\nn \\
%\ea
%\ba
\fft{k^*}{\w^2}&=&1-\fft{(26+20\sqrt[3]{2}+15\sqrt[3]{2^2})
\ep^4\w^4}{720}\nn\\
&&-\fft{(80+63\sqrt[3]{2}+49\sqrt[3]{2^2})\ep^6\w^6}{6048}
%&&-\fft{124+98\sqrt[3]{2}+77\sqrt[3]{2^2}\ep^8\w^8}{51840}
%\,\nn\\
%&&-\fft{10147+8074\sqrt[3]{2}+6435\sqrt[3]{2^2}\w^{10}\ep^{10}}{23950080}\,\nn\\
%&&\left.-\fft{12871+10218\sqrt[3]{2}+8112\sqrt[3]{2^2}
%\w^{12}\ep^{12}}{28304640}
+\cdots.\nn
\ea
The modified Hamiltonian now deviates from the original 
Hamiltonian beginning at the fourth order.

In solving for periodic motion, the most serious error is
not the energy, which is well conserved by symplectic
integrators after each period. The
most serious error is the phase (or angle) error, which
accumulates after every period. The phase error thus grow
linearly with integration time, even for symplectic integrators. 
The phase error is simply related to the fractional deviation of the the
modified angular frequency from the original frequency:
%%%%%
\be
\Delta\phi=(\w_A-\w){\rm T}=2\pi({{\w_A}\over{\w}} -1)\,.\la{phase}
\ee
%%%%%
For an $n$th order algorithm, its phase error is essentially given by 
the coefficient $c_n$,
\be 
{{\w_A}\over{\w}} -1 = {{\theta}\over{\ep\w}}-1=c_n\ep^n\w^n +O(\ep^{n+2})
\la{cff}
\ee
For the SV algorithm, we have from (\ref{waexp}), $c_2=1/24$. 
Below we list the
$c_4$ coefficient for some recent fourth order algorithms.
\ba
c_4(FR)&=& -\fft{(32+25\sqrt[3]{2}
+20\sqrt[3]{2^2})}{1440}\nn\\
       &=&-0.0661431\nn\\
c_4(M)&=& \fft{-2956612 + 124595\sqrt{471}}{2797262640}
\nn\\
       &=& -0.0000902971\nn\\
c_4(BM)&=& -0.0000133432    
\nn\\
c_4(C)&=& \fft1{7680}= 0.000130208
\la{clist}
\ea
FR is the Forest-Ruth algorithm which requires only 3 force
evaluations. M is McLachan's algorithm\cite{mclach} with 4 force 
evaluations. BM is Blanes and Moan's algorithm\cite{blanmoan} with 
6 force evaluations. C is Chin's all positive time steps, forward 
algorithm\cite{chin97} C which uses 3 force and 1 force gradient 
evaluation. Note that coefficient is positive only for 
the forward algorithm. Since for the same number of
force evaluations, one can update the FR algorithm twice
at half the step size as compared to the BM algorithm, one should
reduce FR's error coefficient by a factor $2^4=16$ before comparing
it to BM's error coefficient. Or, one should multiplying BM's error 
by $(6/3)^4$ before comparing to that of FR's. Thus for an equal 
computational effort comparison, we should multiply each algorithm's 
coefficient by $(n/3)^4$ where $n$ is the number of force evaluation 
and normalize it by dividing by FR's value. This normalized 
coefficient $c^*$ for each algorithm is then as follow, (For 
this comparison, we count each force gradient as roughly
equivalent to one force evaluation. For most forces in celestial
mechanics and molecular dynamics, the force gradient can be
evaluate without too much effort.)
\ba
c_4^*(FR)&=&-1.0000\nn\\
c_4^*(M)&=& -0.0043\nn\\
c_4^*(BM)&=& -0.0032\nn\\
c_4^*(C)&=&\,\,\,\, 0.0062
\la{cnorm}
\ea
Since the dynamics of the harmonic oscillator is very simple,
its phase error is only a necessary but not a sufficient
criterion for gauging an algorithm efficiency. While it
is well known that FR has relatively large error among
fourth order algorithms, it is interesting 
that this simple {\it analytical} calculation can indeed 
distinguish the latter three algorithms as significantly better 
than FR. (In the case of the forward algorithm, we have 
deliberately chosen algorithm C to do the comparison. 
Within the parametrized family of fourth order forward 
algorithms\cite{chinchen02,chinchen03}, one can actually 
make $c_4$ vanish. Thus a general fourth order forward 
algorithm can solve the harmonic oscillator to six order, but 
such a comparison would not have been fair without
allowing other algorithms to be fine-tuned.)   
 
\section{Concluding summary}

     In this work, we have shown that the evolution of 
any time-reversible symplectic algorithm can be solved in closed form.
We proved that any factorized, time reversible symplectic algorithm
must have a phase-space matrix of the form (\ref{gentr}), with equal
diagonal elements. 
Based on the exact solution, we have further shown that 1)
the Lie series expansion for the modified Hamiltonian converges
in the case of the harmonic oscillator. The global structure
of the modified Hamiltonian is known. 2) The phase-space structure 
of the harmonic oscillator is closely preserved by time-reversible 
algorithms but is distorted by non-reversible integrators. 
3) The analytical form of the phase error can be used to broadly 
assess the effectiveness of any reversible algorithm without doing any 
numerical calculation. We verified this claim by computing the phase
error coefficients for four fourth-order algorithms.

\begin{acknowledgments}
This work was supported in part, by a National Science Foundation 
grant (to SAC) No. DMS-0310580 and MOS funding from the
Institute for Quantum Studies, Texas A\&M University.
\end{acknowledgments}
%%%%%%%%%%%%%%%%%%%%%%%%%%%%%%%%%%%%%%%%%%%%%%%%%%%%%%%%%%%%%%%%%%%%%%%%%%%%%%%%%
%\newpage
%\bigskip
%\bigskip
%\centerline{REFERENCES}
%\vspace{.3 truein}

%\bibliography {dmts}
%\bibliographystyle{revtex4}

%%%%%%%%%%%%%%%%%%%%%%%%%%%%%%%%%%%%%%%%%%%%%%%%%%%%%%%
\end{document}